\documentclass[floatfix,aps,amsmath,nofootinbib,twocolumn,10pt]{revtex4}

\usepackage{listings}
\usepackage{graphicx}
\usepackage{bm}
\usepackage{rotating}
\usepackage{array}
\usepackage{amsmath}
\usepackage{amssymb} %花体字母加粗
\usepackage{mathrsfs} %花体字母
\usepackage{cancel}
\usepackage{subfig}
\usepackage{float}
\usepackage{caption}
\usepackage[dvipsnames]{xcolor}
\usepackage[normalem]{ulem}

\newcommand{\Mov}[1]{{\color{black}{#1}}}

\def\({\left(}
\def\){\right)}
\def\[{\left[}
\def\]{\right]}

\def\e{\begin{equation}}
\def\q{\end{equation}}
\def\m{\begin{eqnarray}}
\def\n{\end{eqnarray}}

\begin{document}
\title{Model-independent Test of the Cosmic Anisotropy with Inverse Distance Ladder}% Force line breaks with \\
%\thanks{A footnote to the article title}%
\author{Zong-Fan Yang$^{1}$}
\author{Da-Wei Yao$^{2}$}
\author{M. Le Delliou$^{1,3,4,5,6,7}$}
\author{Ke Wang$^{8}$}
\thanks{Corresponding author: {wangkey@lzu.edu.cn}}
\affiliation{$^1$School of Physical Science and Technology, Lanzhou University, Lanzhou 730000, China}
\affiliation{$^2$Beijing Zhihuo Technology Co., Ltd, Beijing 100000, China}
\affiliation{$^3$Institute of Theoretical Physics $\&$ Research Center of Gravitation, Lanzhou University, Lanzhou 730000, China}
\affiliation{$^4$Key Laboratory of Quantum Theory and Applications of MoE, Lanzhou University, Lanzhou 730000, China}
\affiliation{$^5$Lanzhou Center for Theoretical Physics $\&$ Key Laboratory of Theoretical Physics of Gansu Province, Lanzhou University, Lanzhou 730000, China}
\affiliation{$^6$Instituto de Astrof\'isica e Ci\^encias do Espa\c co, Universidade de Lisboa,
Faculdade de Ci\^encias, Ed.~C8, Campo Grande, 1769-016 Lisboa, Portugal}
\affiliation{$^7$Universit\'e de Paris-Cit\'e, APC-Astroparticule et Cosmologie (UMR-CNRS 7164), 
%Batiment Condorcet, 10 rue Alice Domon et L\'eonie Duquet, F-75205 Paris Cedex 13, France.
F-75006 Paris, France}
\affiliation{$^8$Department of Physics, Liaoning Normal University, Dalian 116029, China}

\date{\today}% It is always \today, today,
             %  but any date may be explicitly specified

\begin{abstract}
\Mov{If t}%T
he Universe \Mov{is }endowed with %the 
cosmic anisotropy\Mov{, it} will have a preferred direction of expansion. %Therefore, r
Reconstructing the expansion history by Gaussian Process (GP) can be used to probe the cosmic anisotropy model-independently. In this paper, for the luminosity distance $d_L(z)$ reconstruction, we turn to the inverse distance ladder, where %the 
type Ia supernova (SNIa) from the Pantheon+ sample determine %the 
relative distances and %the 
strongly gravitationally lensed quasars from H0LiCOW sample anchor these relative distances with some absolute distance measurements. 
By isolating the anisotropic information that could be %\Mout{maybe} 
carried by the Hubble constant $H_0$ and obtaining %the 
constraints on the intrinsic parameter of SNIa, the absolute magnitude $M=-19.2522^{+0.0594}_{-0.0649}$ (at $68\%$ CL), we find that $d_L(z)$ reconstructions from samples located in different region of the Galactic coordinate system are almost consistent with each other, while %\Mout{and} 
only a very weak preference for the cosmic anisotropy is found.
\end{abstract} 

%\keywords{Suggested keywords}%Use showkeys class option if keyword
                              %display desired
\maketitle

%\tableofcontents

\section{Introduction}
\label{sec:intro}
The cosmological principle hypothesizes that the Universe features homogeneity and isotropy on large scales. Rooted %\Mout{Ground} 
on some basic important assumptions, including the cosmological principle, the standard model of particle physics and Einstein’s theory of gravity, the standard model of cosmology is established.
More precisely, according to the latest cosmic microwave background (CMB) observations~\cite{Planck:2018vyg}, %the 
measurements of galaxy clustering and weak gravitational lensing~\cite{DES:2021wwk} and %the 
measurements of baryon acoustic oscillation~\cite[BAO]{DES:2024cme,eBOSS:2020yzd,DESI:2024mwx}, the Lambda cold dark matter ($\Lambda$CDM) model is generally considered as being supported.
Despite its successes on large cosmic scales, there are some underlying crises appearing on small scales, such as the presence of a dipole %with larger amplitude \Mov{than the CMB dipole }
in the quasar~\cite{Secrest:2020has,Dam:2022wwh,Kothari:2022bjr,Panwar:2024xum}, radio galaxy~\cite{Secrest:2022uvx,Wagenveld:2023kvi} and active galactic nuclei sky~\cite{Singal:2024ldf}, a smaller than predicted amplitude of matter fluctuations $\sigma_8$ obtained from cluster counts~\cite{Planck:2015lwi} and a larger local value of the Hubble constant $H_0$~\cite{Riess:2021jrx} compared with CMB observations~\cite{Planck:2018vyg}.
Besides modifying general relativity (GR), introducing dynamical dark energy models or different treatments of systematic uncertainty, these tensions also might be alleviated by departing from %challenge 
the cosmological principle.
In particular, the %Especially a 
$5.1\sigma$ tension between the number-count dipoles found jointly in the quasar and radio galaxy sky~\cite{Secrest:2022uvx}, and 
the CMB kinematic dipole~\cite{Planck:2018nkj}, challenges the cosmological principle directly.
Therefore, it is imperative to know to what extent the cosmological principle is violated%\Mout{\Mcom{???}the truth}
, hence the need for a further test of %the 
local cosmic inhomogeneity and anisotropy% on small scales
.

The most simple methods to test %the 
cosmic anisotropy are based on %the 
type Ia supernovae (SNIa) data, such as the Pantheon+ sample~\cite{Brout:2022vxf}. Since a large number of SNIa are spread out in the sky, one straightforward method is to divide the whole SNIa sample into several subsets according to its distribution. In particular, the hemisphere comparison method divides the whole sample into two data subsets and investigates the discrepancy between these two opposite hemispheres~\cite{Schwarz:2007wf,Antoniou:2010gw}. To further probe the fine structure that might be smoothed out, one can use Hierarchical Equal Area iso-Latitude Pixelization~\cite[$\mathsf{HEALPix}$]{Gorski:2004by} to divide the whole SNIa sample into several data subsets at a time. Another common method is the dipole fitting method, which assumes the observation of every SNIa depends on an {\it a priori} cosmic %priori 
dipole~\cite{Mariano:2012wx}. The above three methods propose means of extracting 
cosmic anisotropy from SNIa data. \Mov{However, most common methods further process the SNIa data in order to retrieve anisotropies in specific cosmological parameters, which in turn requires assuming a specific cosmological model.} %In addition, they require to further 
% treat the SNIa data in order to finally put the cosmic anisotropy into some specific parameters. This further treatment can be simply effected 
% by assuming a specific cosmological model. %However, these methods are model-dependent. 
More precisely, %in these methods, 
the luminosity distance $d_L$ is calculated in a specific cosmological model, such as in the $\Lambda$CDM model~\cite{Deng:2018yhb,Sun:2018cha,Deng:2018jrp,Zhao:2019azy,Hu:2023eyf}, $w$CDM model~\cite{Zhao:2013yaa}, CPL model~\cite{Lin:2015rza}, Finslerian model~\cite{Chang:2019utc}, \Mov{or} $\Lambda$LTB model~\cite{Wang:2023reg}% and so on
. However, the introduction of such model is not only not generic enough, but also source of potential biases.

If $d_L$ can be calculated model-independently, one can make a model-independent test of the cosmic anisotropy with SNIa data. In fact, there are several ways to realize this ambition.
\Mov{For example, there are families of polynomials, function of redshift variables~\cite{Semiz:2015gga,Camlibel:2020xbn,Collett:2019hrr}, or using the Pad\'e approximation~\cite{Hu:2024qnx}, in each case approximating $d_L$ model-independently.}
Similarly, $d_L$ can be expressed model-independently by the cosmographic parameters according to the well-known cosmographic approach~\cite{Andrade:2018eta,Cattoen:2007id}. 
As for the completely data-driven nonparametric approach, $d_L$ can be reconstructed model-independently by %the 
artificial neural network \cite[ANN]{Wang:2019vxv}.
In some cases, one may only care about the relative distances from SNIa, where one can turn to Gaussian Process (GP) to reconstruct the dimensionless luminosity distance $D_L\equiv d_LH_0/c$~\cite{Jesus:2019nnk,Seikel:2012uu}.
In this paper, since we start from $D_L$, as GP proved useful to treat data model-independently, we will use GP to perform the reconstruction of $D_L$ from the Pantheon+ sample~\cite{Brout:2022vxf}.

Given the relative distances accessible from SNIa, one should anchor them with an absolute distance measurement.
For the local distance ladder, $H_0$ and the absolute magnitude $M$ of SNIa are usually calibrated by very local Cepheids~\cite{Brout:2022vxf,Riess:2021jrx}. Although these very local absolute distance measurements are almost completely insensitive to the cosmological background model, they are too local to consider their spatial distribution, hence no anisotropic information can be extracted from them.
On the contrary, the inverse distance ladder can be calibrated at high redshift using other independent measurements, such as the sound horizon at radiation drag $r_d$ determined by CMB observations~\cite{BOSS:2014hhw,Cuesta:2014asa}, the time delays $\Delta t_{j,1}$ between the Refsdal Supernova% Refsdal
’s $j$th and first appearances~\cite{Kelly:2023mgv,Li:2024elb}, \Mov{or} the time-delay distance $d_{\Delta t}$ of strongly gravitationally lensed quasars~\cite{Collett:2019hrr,Wong:2019kwg,Liao:2019qoc,Taubenberger:2019qna,Liao:2020zko,Li:2023gpp}% and son on
. 
If anchors at high redshift are also spread out in the sky, they may carry certain anisotropic information, just as SNIa samples do% dose
.
Therefore, in this paper, we will use $d_{\Delta t}$ measurements of six lensed quasars from the H0LiCOW sample~\cite{Wong:2019kwg} to anchor our $D_L$ reconstructions from the Pantheon+ sample~\cite{Brout:2022vxf}.
That is to say, we can model-independently test the cosmic relative or absolute anisotropy with $D_L$ reconstructions only or the combination of $D_L$ reconstructions and $d_{\Delta t}$ measurements respectively.

\Mov{Should there exist cosmic anisotropy, the anisotropic information should affect all observations. 
Therefore a consistency test between different observations can further confirm the anisotropic information. On one hand, as the spatial separation between Cepheids %is not wide enough
restricts due to their small distances, they are not producing measurable anisotropy effects to cross-check with the Pantheon+ sample~\cite{Brout:2022vxf}. The H0LiCOW sample~\cite{Wong:2019kwg}, on the other hand, consisting of objects located at higher redshifts, not only can serve as anchor of the inverse distance ladder, but also can carry the anisotropic information, which can then be cross-checked with the Pantheon+ sample. B%That is to say, b
y choosing the inverse ladder anchor, we can extract anisotropies from the combined set of %whole 
data% combination
, beyond the limits of %not just 
SNIa data. As a result of the inverse ladder approach, the consistency test in the latter part of our paper brings an advantage over the usual distance ladder. A successful test would reveal the presence of cosmic anisotropy in the entire data combination; a failed test proves that the anisotropy which appears in the Pantheon+ sample, but disappears in H0LiCOW sample, probably results from potential systematic uncertainties.}

This paper is organized as follows.
In section~\ref{sec:MD}, we introduce a model-independent method and the widespread data through the sky %\Mout{Galactic coordinate system} 
for $D_L(z)$ reconstruction, namely GP and the Pantheon+ sample~\cite{Brout:2022vxf}. \Mov{Then}%Meanwhile
, we present the anchors from the H0LiCOW sample~\cite{Wong:2019kwg} for the inverse distance ladder.
In section~\ref{sec:R}, we show the reconstruction of $d_L(z)$ from different datasets which serves as %the 
probe of the cosmic anisotropy in our paper.
Finally, a brief summary and discussions are included in section~\ref{sec:SD}.

\section{Methodology and data}
\label{sec:MD}
In this section, we provide a brief summary of the model-independent GP algorithm, the SNIa data used for the Universe expansion history reconstruction and the strong gravitational lenses data used to anchor the inverse distance ladder.
\subsection{Gaussian Process}
\label{ssec:gp}
While the Gaussian distribution describes the distribution of a random variable with mean and variance, GP describes the distribution of a function $f(x)$ with mean $\mu(x)$, variance $\sigma^2(x)$ and kernel $k(x,x')$. There are several choices of kernel function: %, such as the 
squared exponential kernel, Matern $(5/2)$ kernel, Matern $(7/2)$ kernel \Mov{or }%and 
Matern $(9/2)$ kernel\Mov{, among others}. Here we only use the squared exponential kernel
\begin{equation}
k(x,x')=\sigma_f^2 \exp{\left[-\frac{(x - x')^2}{2l^2}\right]},
\end{equation}
where $\sigma_f$ and $l$ are the GP hyperparameters. The former %one 
determines the correlation strength between $f(x)$ and $f(x')$ at two different points, the latter gives %one is 
the correlation length determining the range of such correlation%which $f(x')$ should be correlated with $f(x)$
.

Before reconstructing $f(x)$% is reconstructed
, these two hyperparameters should be known. Given the observed data $\{\bm{x}, \bm{y}=f(\bm{x})\pm\sigma(\bm{x})$\}, $\sigma_f$ and $l$ can be determined %\Mout{trained} 
by maximizing the marginal likelihood~\cite{Seikel:2012uu,Shafieloo:2012ht}
\begin{align}
\ln \mathcal{L} = & \ln P(\bm{y} | \bm{x}, \sigma_f, l) \\ \nonumber
=&-\frac{1}{2} (\bm{y} - \mu(\bm{x}))^T[k(\bm{x}, \bm{x}) + C]^{-1} (\bm{y} -\mu(\bm{x})) \\ \nonumber
&- \frac{1}{2} \ln |k(\bm{x}, \bm{x}) + C| - \frac{n}{2} \ln 2\pi,
\end{align}
where $C=C(\sigma(\bm{x}),\sigma(\bm{x}))$ is the covariance matrix of the data and $n$ is the number of data points.
After optimizing for $\sigma_f$ and $l$, $f(x)$ at points ${\bm x}^*$ can be reconstructed as the mean of $f(\bm{x}^{*})$
\begin{equation}
\overline{f(\bm{x}^{*})} = \mu(\bm{x}^{*}) + k(\bm{x}^{*}, \bm{x})[k(\bm{x}, \bm{x}) + C]^{-1}(\bm{y} - \mu(\bm{x}))
\end{equation}
with the covariance of $f(\bm{x}^{*})$
\begin{align}
{\rm cov}(f(\bm{x}^{*}))= k(\bm{x}^{*}, \bm{x}^{*})-k(\bm{x}^{*}, \bm{x})[k(\bm{x}, \bm{x}) + C]^{-1}k(\bm{x}, \bm{x}^{*}).
\end{align}
Following the above procedures, we can reconstruct the cosmic expansion history by the GP package $\mathtt{GaPP3}$~\cite{Seikel:2012uu}.

\subsection{Pantheon+ sample}
\label{ssec:p}
\begin{figure*}[]
\begin{center}
\subfloat{\includegraphics[width=0.8\textwidth]{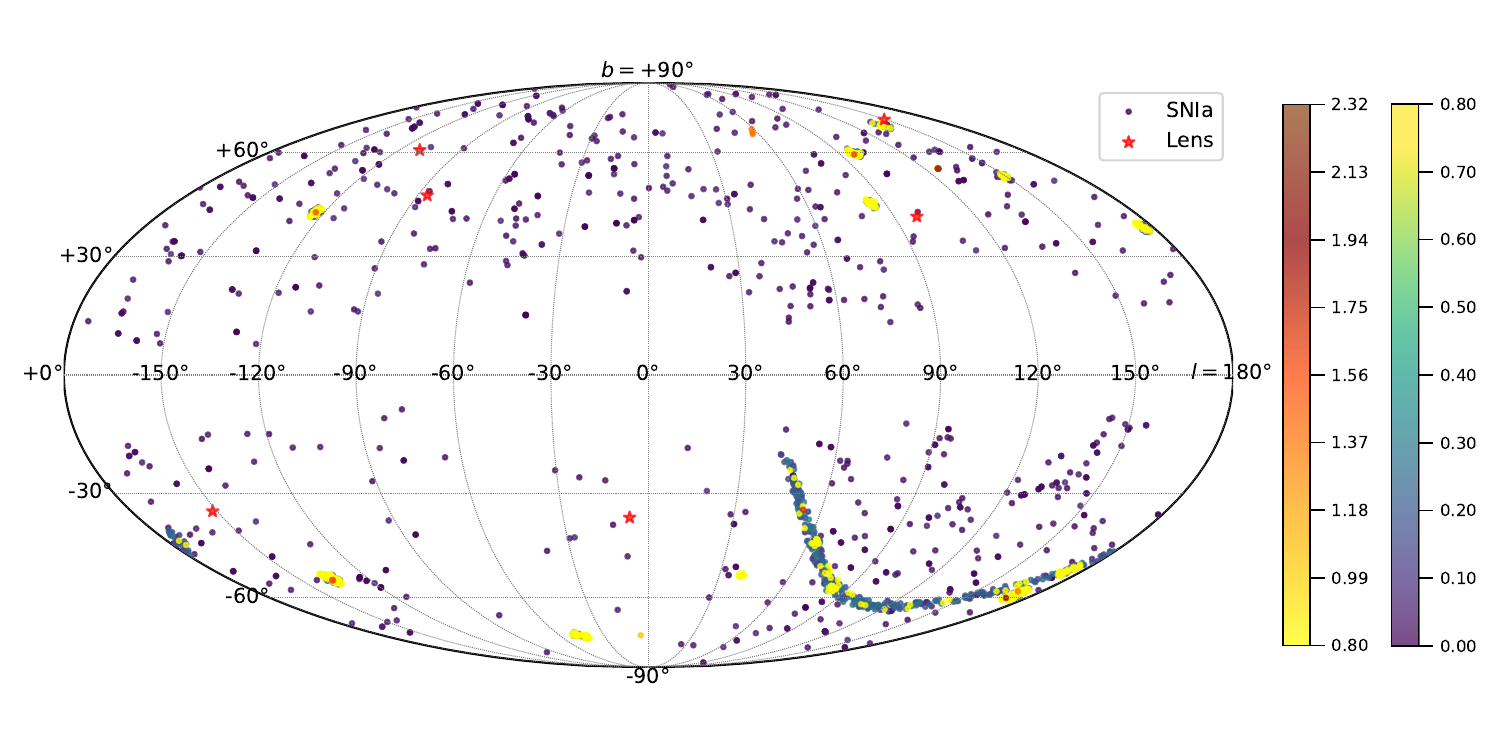}}
\end{center}
\captionsetup{justification=raggedright}
\caption{D%The d
istribution of 1550 SNIa of the Pantheon+ sample~\cite{Brout:2022vxf} and 6 lenses of the H0LiCOW sample~\cite{Wong:2019kwg} in the Galactic coordinate system $(l, b)$, where the %blue
coloured dots represent SNIa with the \Mov{two }side bars indicating their redshifts % dots are SNIa 
and the red stars are lensed quasars.}
\label{fig:data}
\end{figure*}

In this paper, we only use the Pantheon+ sample~\cite{Brout:2022vxf} to reconstruct $D_L(z)$. This dataset consists of $1701$ light curves of $1550$ distinct SNIa in the redshift range $0.001<z<2.26$.
%Here we need to transform the distribution of SNIa from the Equatorial coordinate system $(\alpha, \delta)$ into the Galactic coordinate system $(l, b)$\Mov{???}. 
In Fig.~\ref{fig:data}, we show the distribution of these $1550$ SNIa in the sky %\Mout{Galactic coordinate system $(l, b)$}. 
Following the hemisphere comparison method~\cite{Schwarz:2007wf,Antoniou:2010gw}, we divide the full dataset into several subsets to probe %the 
cosmic anisotropy, as listed in Tab.~\ref{tb:sample}. Each %Obviously, each 
SNIa subset provides %a simplified view on spatial 
%can only give the local 
information on a corresponding patch of the Universe.
\begin{table*}
\captionsetup{justification=raggedright}
\caption{Definitions of the full sky spatial divisions in the Galactic coordinate system $(l,b)$, resulting in corresponding subsamples from %According to 
the distribution of 1550 SNIa of the Pantheon+ sample~\cite{Brout:2022vxf} and 6 lenses of the H0LiCOW sample~\cite{Wong:2019kwg}. The % in the Galactic coordinate system $(l,b)$, the 
full dataset is divided into $\rm{PH_{ N}}$+$\rm{PH_{S}}$ by a split %division 
along the Galactic plane, $\rm{PH_{E}}$+$\rm{PH_{W}}$ by a division along a plane orthogonal to the Galactic plane, $\rm{PH_{NE}}$+$\rm{PH_{NW}}$+$\rm{PH_{SE}}$+$\rm{PH_{SW}}$ by both simultaneous divisions and $\rm{PH_{U}}$+$\rm{PH_{D}}$ by a division along a plane orthogonal to the CMB dipole ($l=264.02$°$\pm0.01$°$,b=48.25$°$\pm0.01$°)~\cite{Planck:2018nkj}.}
\label{tb:sample}
\begin{tabular}{cccc}
\hline
\hline
\rm Dataset & $l$ &  $b$ & \Mov{Size}\\
\hline
$\rm{PH_{All}}$ &  -180°$\leq l < $180° &  $-90$°$\leq b \leq $90° & $1701$  \\
\hline
$\rm{PH_{N}}$ &  -180°$\leq l < $180° &  0°$\leq b \leq $90°  & $664$    \\
$\rm{PH_{S}}$ &  -180°$\leq l < $180° &  $-90$°$\leq b <$0°  & $1037$    \\
\hline
$\rm{PH_{W}}$  &  -180°$\leq l < $0°&  $-90$°$\leq b \leq $90°  & $549$    \\
$\rm{PH_{E}}$  &  0°$\leq l < $180°  &  $-90$°$\leq b \leq $90°  & $1152$   \\
\hline
$\rm{PH_{NW}}$  &   -180°$\leq l < $0° &  0°$\leq b \leq $90°  & $303$    \\
$\rm{PH_{NE}}$  &  0°$\leq l < $180° &  0°$\leq b \leq $90°  & $361$    \\
$\rm{PH_{SW}}$  &  -180°$\leq l < $0° &  $-90$°$\leq b <$0°  & $246$    \\
$\rm{PH_{SE}}$  &  0°$\leq l < $180° &  $-90$°$\leq b <$0°  & $791$    \\
\hline
$\rm{PH_{U}}$  &  $(\cos b\cos l,\cos b\sin l, \sin b)$ & $\cdot~(\cos 48.25\cos 264.02,\cos  48.25\sin 264.02, \sin  48.25)\geq0$  & $634$   \\
$\rm{PH_{D}}$  &  $(\cos b\cos l,\cos b\sin l, \sin b)$ & $\cdot~(\cos 48.25\cos 264.02,\cos  48.25\sin 264.02, \sin  48.25)<0$ & $1067$ \\
\hline 
\hline
\end{tabular}
\end{table*}

Given the corrected apparent magnitudes $m$ of SNIa, %the 
information about the luminosity distance $d_L$ of SNIa can be derived from the relation between them as
\begin{align}
\label{eq:dL}
m&=5\log_{10}\left[\frac{d_L}{1{\rm Mpc}}\right]+25+M\\ \nonumber
&=5\log_{10}\left[\frac{d_LH_0/c}{1{\rm Mpc}\cdot H_0/c}\right]+25+M\\ \nonumber
&=5\log_{10}\left[d_LH_0/c\right]-5\log_{10}\left[10{\rm pc}\cdot H_0/c\right]+M, 
\end{align}
where $M$ is the absolute magnitude of SNIa. By defining a dimensionless luminosity distance $D_L\equiv d_LH_0/c$ and a fiducial parameter $M_*\equiv M-5\log_{10}\left[10{\rm pc}\cdot H_0/c\right]$, we have
\begin{equation}
\label{eq:m2DL}
D_L=10^{\frac{m-M_*}{5}}.   
\end{equation}
It is worth noting that we, in fact, don't know the value of $M_*$% in fact
.
However we can express it as 
\begin{align}
M_*&=-19.3+\delta_M-5\log_{10}\left[\delta_{70}\cdot70/(3\times10^{10})\right]\\\nonumber
&=23.86+\delta_M-5\log_{10}\left[\delta_{70}\right],
\end{align}
where $\delta_M-19.3=M$ and $\delta_{70}\cdot 70~{\rm km/s/Mpc}=H_0$ characterize the unknown information about $M$ and $H_0$. Since $M$ is the intrinsic parameter of SNIa, $\delta M$ (or $M$) should not be space-dependent or time-dependent when Newton’s Constant $G$ is supposed to be truly a constant, remaining unchanged%a real constant and not changing
~\cite{Zhao:2018gwk,Wang:2020bjk,Liu:2024vlt}. 
By isolating the anisotropic information carried by $\delta_{70}$ (or $H_0$), we can determine $\delta_M$ from the following relation
\begin{equation}
\label{eq:m2DLt}
\tilde{D}_L=D_L/\delta_{70}=d_L/c\cdot 70~{\rm km/s/Mpc}=10^{\frac{m-23.86-\delta_M}{5}}.   
\end{equation}
If $\delta_M$ and $\delta_{70}$ are known, according to Eq.~(\ref{eq:m2DL}), we can obtain the covariance matrix of the $\bm{D}_L$ vector of derived $D_L$s from the covariance matrix of the $\bm{m}$ vector as
\begin{equation}
\bm{C}_{D_L}=\bm{J}\bm{C}_m\bm{J}^T,
\end{equation}
where $J_{ii}=\frac{\partial D_{L,i}}{\partial m_i}$ is a diagonal Jacobian matrix. Then we can reconstruct $D_L(z)$ with the derived data $\{\bm{z}, \bm{D}_L, \bm{C}_{D_L}\}$ from the Pantheon+ sample~\cite{Brout:2022vxf}.
Similarly, if only $\delta_M$ is known, according to Eq.~(\ref{eq:m2DLt}), we can reconstruct $\tilde{D}_L(z)$ with the derived data $\{\bm{z}, \bm{\tilde{D}}_L, \bm{C}_{\tilde{D}_L}\}$ from the Pantheon+ sample~\cite{Brout:2022vxf}.

\subsection{H0LiCOW sample}
\label{ssec:l}
\begin{table}
\captionsetup{justification=raggedright}
\caption{Six lensed quasars of the H0LiCOW sample~\cite{Wong:2019kwg} in the Galactic coordinate system $(l, b)$, where $z_l$ and $z_s$ are %is 
the redshifts of lens and source respectively.}
\label{tb:Hsample}
\begin{tabular}{ccccc}
\hline
\hline
Lens name & $l$ &  $b$ & $z_l$ & $z_s$\\
\hline
B$1608+656$&  98.3397 &  40.8915 & $0.6304$ & $1.394$ \\
RXJ$1131-1231$&  -86.2504 &  46.8456  & $0.295$ & $0.654$   \\
HE $0435-1223$&  -151.5704 &  -34.8188  & $0.4546$ & $1.693$   \\
SDSS $1206+4332$&  148.9913  &  71.2445   & $0.745$ & $1.789$  \\
WFI$2033-4723$&  -6.6015  &  -36.5264  & $0.6575$ & $1.662$  \\
PG $1115+080$&  -110.1126 &  60.6443  & $0.311$ & $1.722$   \\
\hline 
\hline
\end{tabular}
\end{table}
The H0LiCOW sample consists of six lensed quasars~\cite{Wong:2019kwg}, as listed in Tab.~\ref{tb:Hsample}. They also spread out in the sky%\Mout{Galactic coordinate system}
, as shown in Fig.~\ref{fig:data}. For cosmological purposes, the effective time-delay distance $d_{\Delta t}$ of every lens system can be considered as %the 
derived data. Because we want to use $d_{\Delta t}$ measurements from the H0LiCOW sample~\cite{Wong:2019kwg} to anchor $D_L(z)$ (or $\tilde{D}_L(z)$) reconstructions at high redshifts, %. So 
we need to convert the dimensionless luminosity distance reconstruction $D_L(z)$ into %the 
dimensionless angular diameter distance reconstruction $D_A(z)=D_L(z)/(1+z)^2$ and consequently into %the 
dimensionless time-delay distance of any lens system 
\begin{align}
\label{eq:Dt}
D_{\Delta t}&\equiv \frac{(1+z_l)D_A(z_l)(1+z_s)D_A(z_s)}{(1+z_s)D_A(z_s)-(1+z_l)D_A(z_l)}=d_{\Delta t}H_0/c,
\end{align}
where $z_l$ and $z_s$ are the redshifts of lens and source respectively.
To isolate the anisotropic information carried by $H_0$, we turn to $\delta_{70}$ again 
\begin{equation}
\label{eq:Dtt}
\tilde{D}_{\Delta t}=D_{\Delta t}/\delta_{70}=d_{\Delta t}/c\cdot 70~{\rm km/s/Mpc}.
\end{equation}

\section{RESULTS}
\label{sec:R}
In this section, we %\Mout{will} 
first determine the space-independent intrinsic parameters of SNIa, $\delta_M$ and $M$. Then we prove that the likelihood that we derive below in Eq.~\eqref{eq:Ld70} does not depend on $\delta_{70}$, nor on % and 
$H_0$% are canceled in the likelihood
. Finally we obtain the space-dependent information from the reconstruction of $d_L(z)$ with known $\delta_M$ and $M$.
\subsection{Determination of $\delta_M$ and $M$}
\label{ssec:dM}
\begin{figure*}[]
\begin{center}
\subfloat{\includegraphics[width=0.43\textwidth]{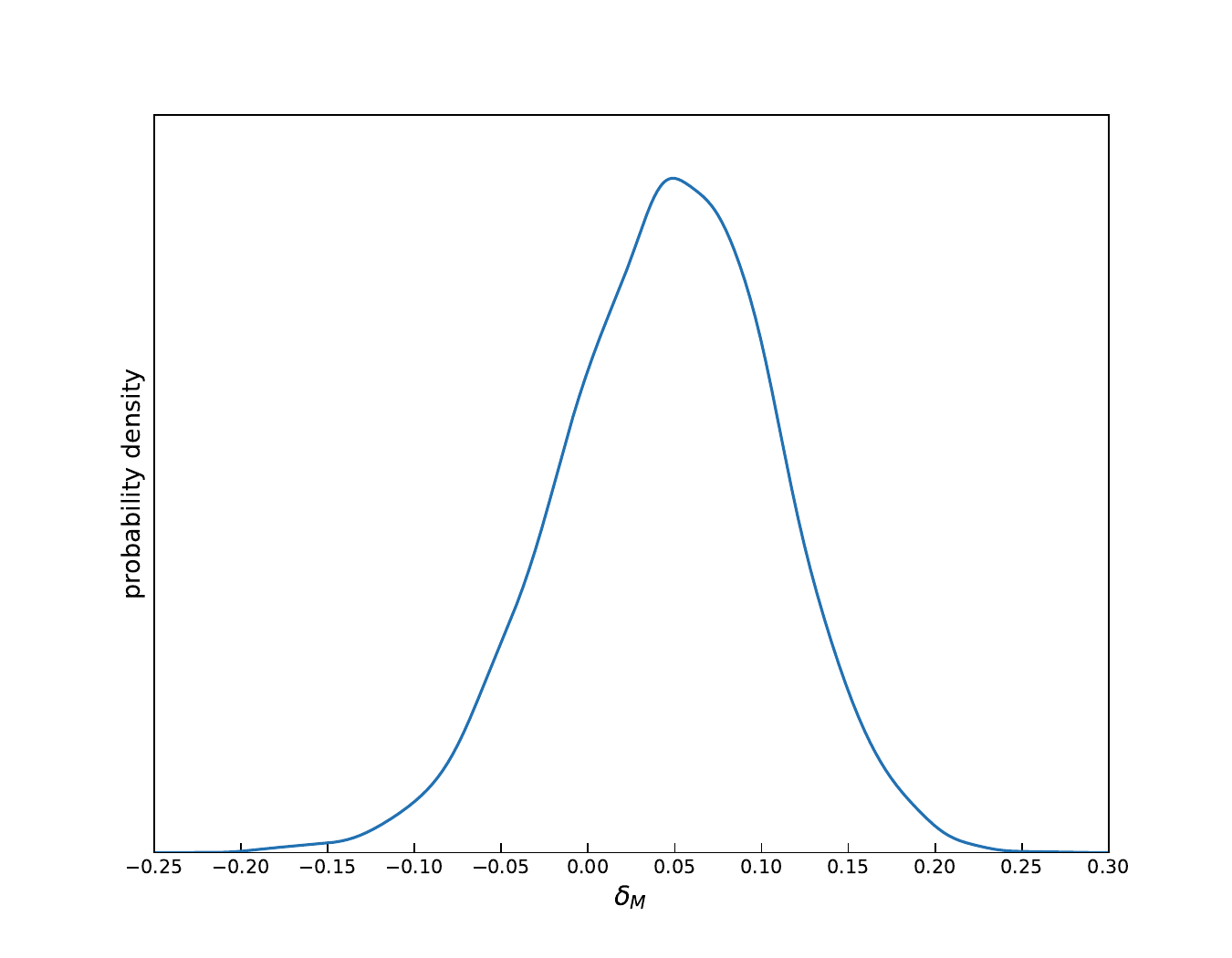}}
\subfloat{\includegraphics[width=0.5\textwidth]{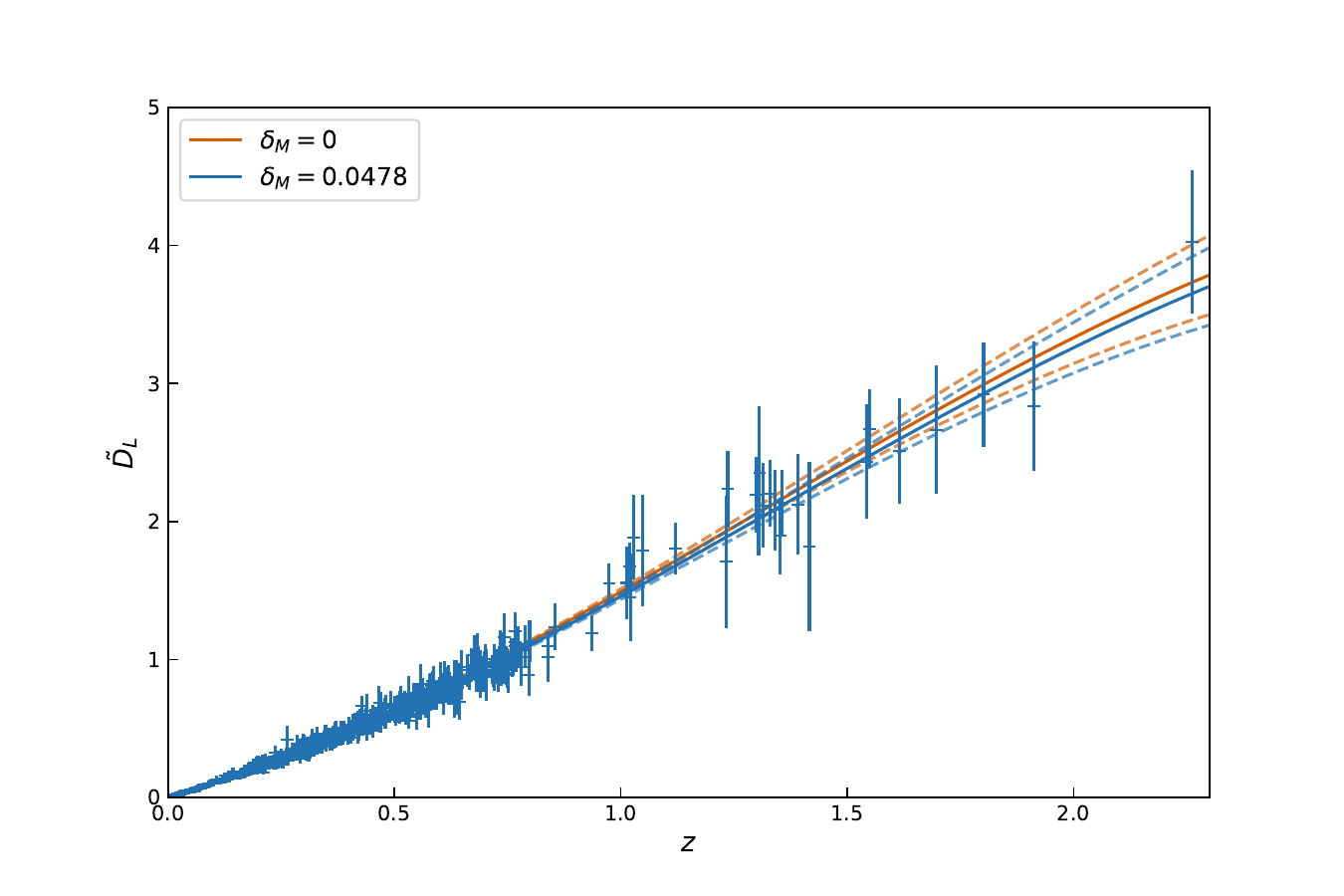}}
\end{center}
\captionsetup{justification=raggedright}
\caption{LEFT: P%The p
osterior PDF of $\delta_M$. RIGHT: Reconstruction of $\tilde{D}_L(z, \delta_M=0)$ (red curve, $\sigma_f=3.0579$, $l=2.4832$) and $\tilde{D}_L(z, \delta_M=0.0478)$ (blue curve, $\sigma_f=2.9913$, $l=2.4832$) from the whole Pantheon+ sample~\cite{Brout:2022vxf}, blue points with errorbars for the $\delta_M=0.0478$ reconstruction, according to Eq.~\eqref{eq:m2DLt} by the GP package $\mathtt{GaPP3}$~\cite{Seikel:2012uu}, where the bracketing dashed curves define %shaded regions are 
the $68\%$ CL regions of the reconstructions.}
\label{fig:all_dl}
\end{figure*}

Before testing %the 
cosmic anisotropy, we must have a good knowledge of the space-independent parameters, such as $\delta_M$ and $M$. 
From %According to 
Eq.~(\ref{eq:Dtt}), the likelihood is
\begin{align}
\label{eq:LdM}
\mathcal{L}&=P(\bm{\tilde{D}}_{\Delta t}, \bm{d}_{\Delta t}|\delta_M,\mathcal{A})\nonumber \\
&=\exp\left[-\frac{1}{2}\left(\frac{c\bm{\tilde{D}}_{\Delta t}(\delta_M)/\bm{d}_{\Delta t}}{70~{\rm km/s/Mpc}}-1\right)^T\bm{C}^{-1}_{\delta_M}\right.\nonumber \\
&\left.~~~~~~~~~~\left(\frac{c\bm{\tilde{D}}_{\Delta t}(\delta_M)/\bm{d}_{\Delta t}}{70~{\rm km/s/Mpc}}-1\right)\right],
\end{align}
where the prior assumption %\Mout{hypothesis} 
$\mathcal{A}$ sets %is that 
the mean value of $c\bm{\tilde{D}}_{\Delta t}/\bm{d}_{\Delta t}$ to %is 
$70~{\rm km/s/Mpc}$, the data vector $\bm{\tilde{D}}_{\Delta t}$ is derived from $\tilde{D}_L(z)$ given the redshift tensor $(\bm{z}_l,\bm{z}_s)$ listed in Tab.~\ref{tb:Hsample}, the corresponding data vector $\bm{d}_{\Delta t}$ is given by the H0LiCOW sample~\cite{Wong:2019kwg} directly and $\bm{C}_{\delta_M}$ is a diagonal matrix proportional to %of 
the variance of $\frac{P(\bm{\tilde{D}}_{\Delta t}|\delta_M)}{P(\bm{d}_{\Delta t})}$.
We use the $\mathtt{emcee}$~\cite{Foreman-Mackey:2012any} sampler to produce %get 
the Monte Carlo Markov Chain (MCMC) of $\delta_M$ with %which has 
a uniform prior distribution over %distribution prior of 
$(-0.3,0.3)$.
During the course of the calculation, each sample of the MCMC chain of $\delta_M$ is used to reconstruct $\tilde{D}_L(z)$ from the whole Pantheon+ sample~\cite{Brout:2022vxf} according to Eq.~(\ref{eq:m2DLt}) by the GP package $\mathtt{GaPP3}$~\cite{Seikel:2012uu}. Due to the prohibitive computational cost of $\tilde{D}_L(z)$ reconstructions with a large number of $\delta_M$, we have to turn to the approximation of 
\begin{equation}
\tilde{D}_L(z, \delta_M\neq0)\approx \tilde{D}_L(z, \delta_M=0)10^{\frac{-\delta_M}{5}}.
\label{eq:app}
\end{equation}
That is to say, we just reconstruct $\tilde{D}_L(z, \delta_M=0)$ once. Then we obtain $\tilde{D}_L(z, \delta_M\neq0)$ %reconstructions 
by adjusting the reconstruction of 
$\tilde{D}_L(z, \delta_M=0)$ with $10^{\frac{-\delta_M}{5}}$, including its %the 
mean value and %the 
uncertainty. Finally, we get the posterior probability density function (PDF) of $\delta_M=0.0478^{+0.0594}_{-0.0649}$ (at $68\%$ CL), as shown in the left panel of Fig.~\ref{fig:all_dl}. 
We also display %Also we show 
the reconstruction of $\tilde{D}_L(z, \delta_M=0.0478)$ in the right panel of Fig.~\ref{fig:all_dl}. We %can 
find that $\delta_M>0$ suppresses $\tilde{D}_L(z)$ and narrows its uncertainty, which agrees with the above approximation, Eq.~(\ref{eq:app}). 

\subsection{Cancellation of $\delta_{70}$ and $H_0$}
\label{ssec:dH}
Given the space-independent parameter evaluation at $\delta_M=0.0478^{+0.0594}_{-0.0649}$ (at $68\%$ CL), we can rewrite Eq.~(\ref{eq:m2DL}) as
\begin{equation}
D_L=10^{\frac{m-23.9078}{5}}\delta_{70}.   
\end{equation}
We then %Then we 
try to determine the space-dependent parameters, such as $\delta_{70}$ and $H_0$.
From %According to 
Eq.~(\ref{eq:Dt}), the likelihood is
\begin{align}
\label{eq:Ld70}
\mathcal{L}&=P(\bm{D}_{\Delta t}, \bm{d}_{\Delta t}|\delta_{70},\mathcal{A})\nonumber \\
&=\exp\left[-\frac{1}{2}\left(\frac{c\bm{D}_{\Delta t}(\delta_{70})/\bm{d}_{\Delta t}}{H_0}-1\right)^T\bm{C}^{-1}_{\delta_{70}}\right.\nonumber \\
&\left.~~~~~~~~~~\left(\frac{c\bm{D}_{\Delta t}(\delta_{70})/\bm{d}_{\Delta t}}{H_0}-1\right)\right],
\end{align}
where the prior assumption %\Mout{hypothesis} 
$\mathcal{A}$ sets %is that 
the mean value of $c\bm{D}_{\Delta t}/\bm{d}_{\Delta t}$ to %is 
$H_0$, the data vector $\bm{D}_{\Delta t}$ is derived from $D_L(z)$, given the redshift tensor $(\bm{z}_l,\bm{z}_s)$ listed in Tab.~\ref{tb:Hsample}, and the corresponding data vector $\bm{d}_{\Delta t}$ is given by the H0LiCOW sample~\cite{Wong:2019kwg} directly and $\bm{C}_{\delta_{70}}$ is a diagonal matrix  proportional to the variance of $\frac{P(\bm{D}_{\Delta t}|\delta_{70})}{P(\bm{d}_{\Delta t})}$.
When we turn to the approximation that%of
\begin{equation}
D_L(z, \delta_{70}\neq1)\approx D_L(z, \delta_{70}=1)\delta_{70},
\end{equation}
we just need to reconstruct $D_L(z, \delta_{70}=1)$ once. Then we obtain $D_L(z, \delta_{70}\neq1)$ %reconstructions 
by adjusting the reconstruction of $D_L(z, \delta_{70}=1)$ with $\delta_{70}$, including its %the 
mean value and %the 
uncertainty.
 Consequently, $\delta_{70}$ cancels out from the prior assumption $\mathcal{A}$, as %\Mout{is canceled in the hypothesis $\Mov{\mathcal{A}}$ because}%\begin{align*}
% \frac{c\bm{D}_{\Delta t}}{\bm{d}_{\Delta t}}= & H_{0}
% \end{align*} and therefore \begin{align*}
% \frac{c\bm{D}_{\Delta t}(\delta_{70}=1)}{\bm{d}_{\Delta t}}= & 70~{\rm km/s/Mpc}
% \end{align*} 
$c\bm{D}_{\Delta t}/\bm{d}_{\Delta t}=H_0$ therefore %\Mout{or} 
$c\bm{D}_{\Delta t}(\delta_{70}=1)/\bm{d}_{\Delta t}= 70~{\rm km/s/Mpc}$, which is independent from $\delta_{70}$. That is to say, the likelihood Eq.~(\ref{eq:Ld70}) can not constrain %'t give the constraints on 
$\delta_{70}$, and thus % and 
$H_0$.

\subsection{Reconstruction of $d_L(z)$}
\label{ssec:dL}
\begin{figure*}[]
\begin{center}
\subfloat{\includegraphics[width=0.5\textwidth]{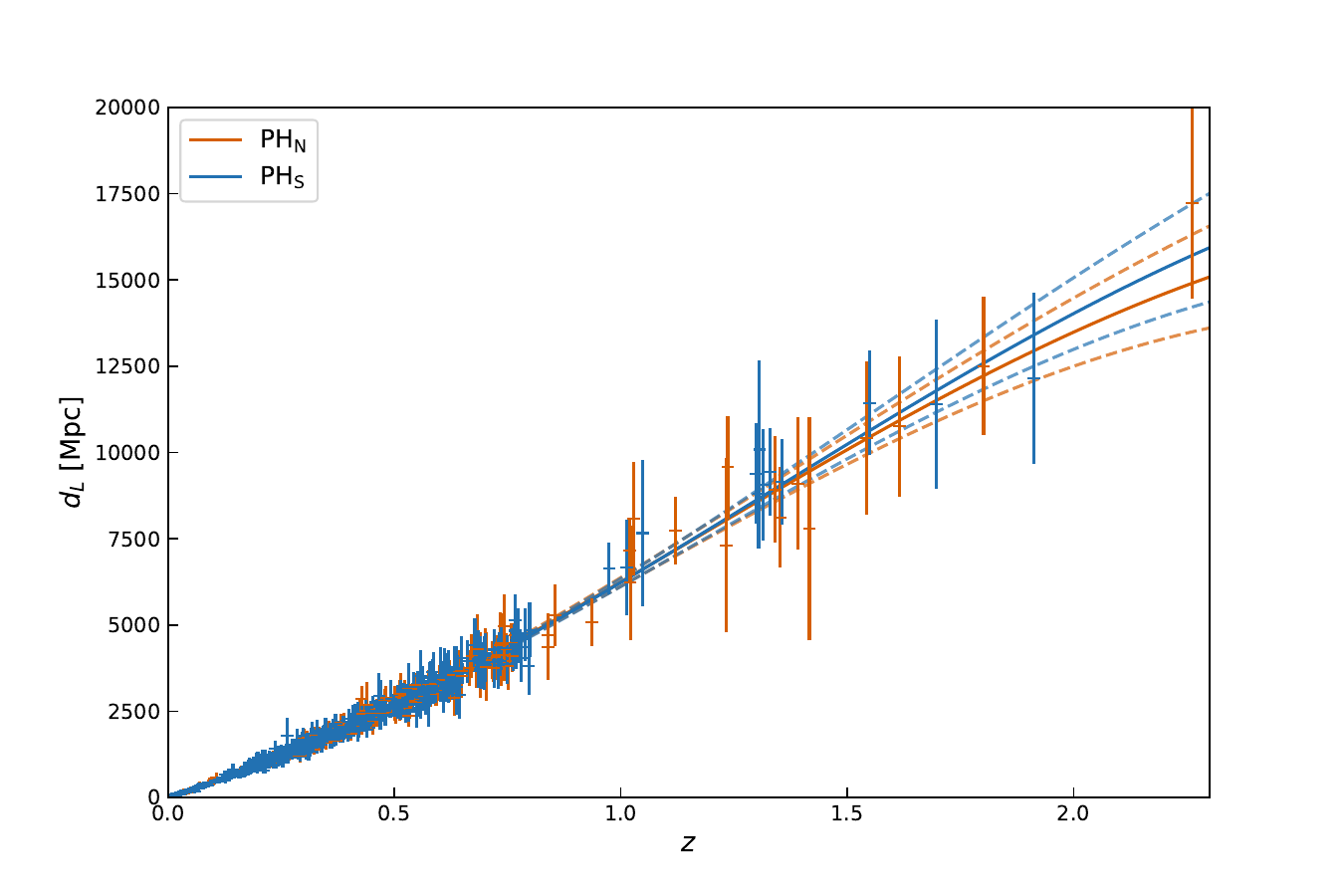}}
\subfloat{\includegraphics[width=0.5\textwidth]{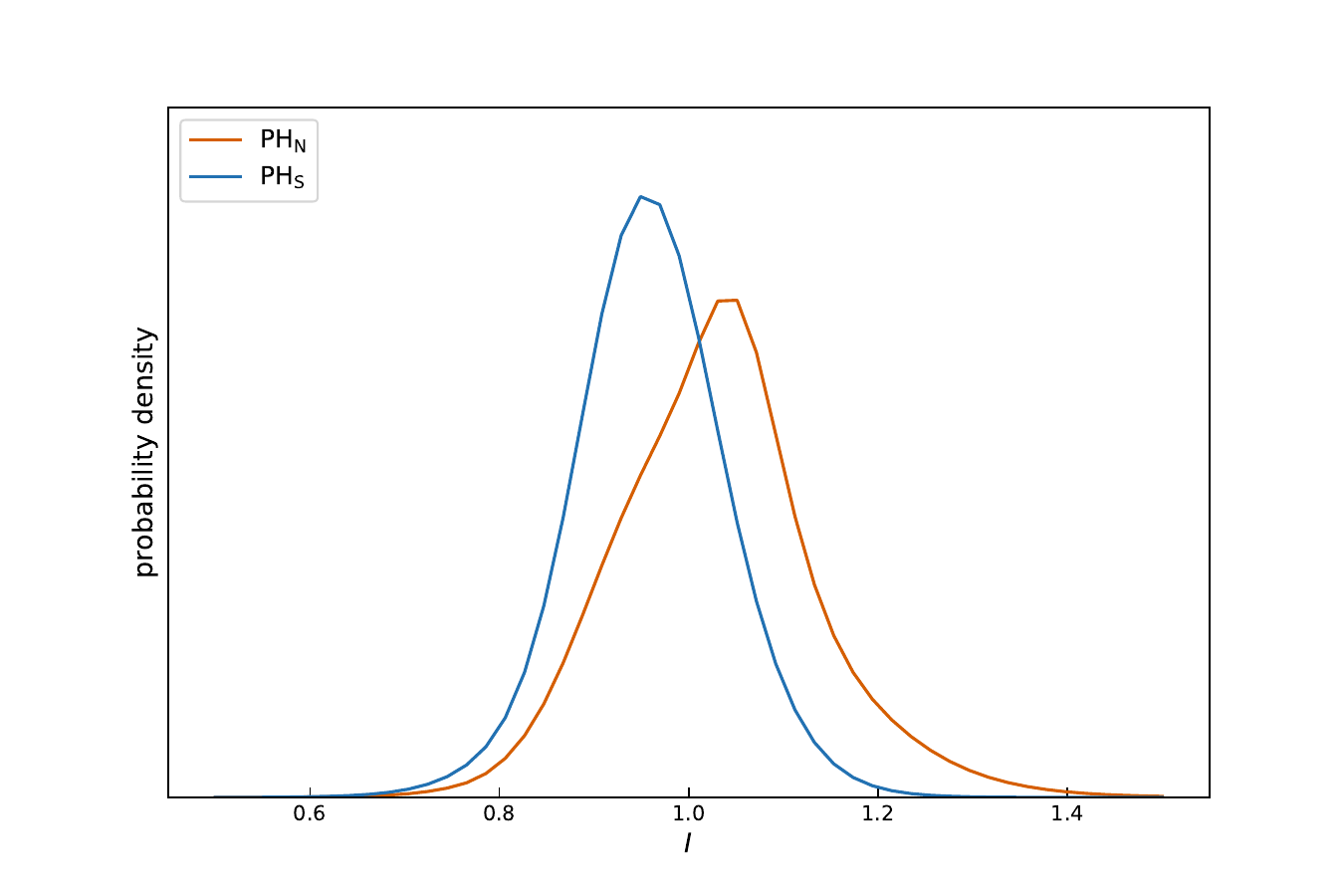}}
\end{center}
\captionsetup{justification=raggedright}
\caption{LEFT: Reconstruction of $d_L(z, \delta_M=0.0478)$ from the $\rm{PH_{N}}$ %dataset (red) 
and $\rm{PH_{S}}$ datasets (red, $\sigma_f=1.1322 \times 10^{4}$~Mpc, $l=2.2562$, and blue, $\sigma_f=1.2099 \times 10^{4}$~Mpc, $l=2.3741$, resp.)% (blue)
, where the bracketing dashed curves define %shaded regions are 
the $68\%$ CL regions of the reconstructions. RIGHT: Constraints on $I$ from the $\rm{PH_{N}}$ %dataset (red) 
and $\rm{PH_{S}}$ datasets (red and blue, resp.)% (blue)
.}
\label{fig:ns_dl}
\end{figure*}

\begin{figure*}[]
\begin{center}
\subfloat{\includegraphics[width=0.5\textwidth]{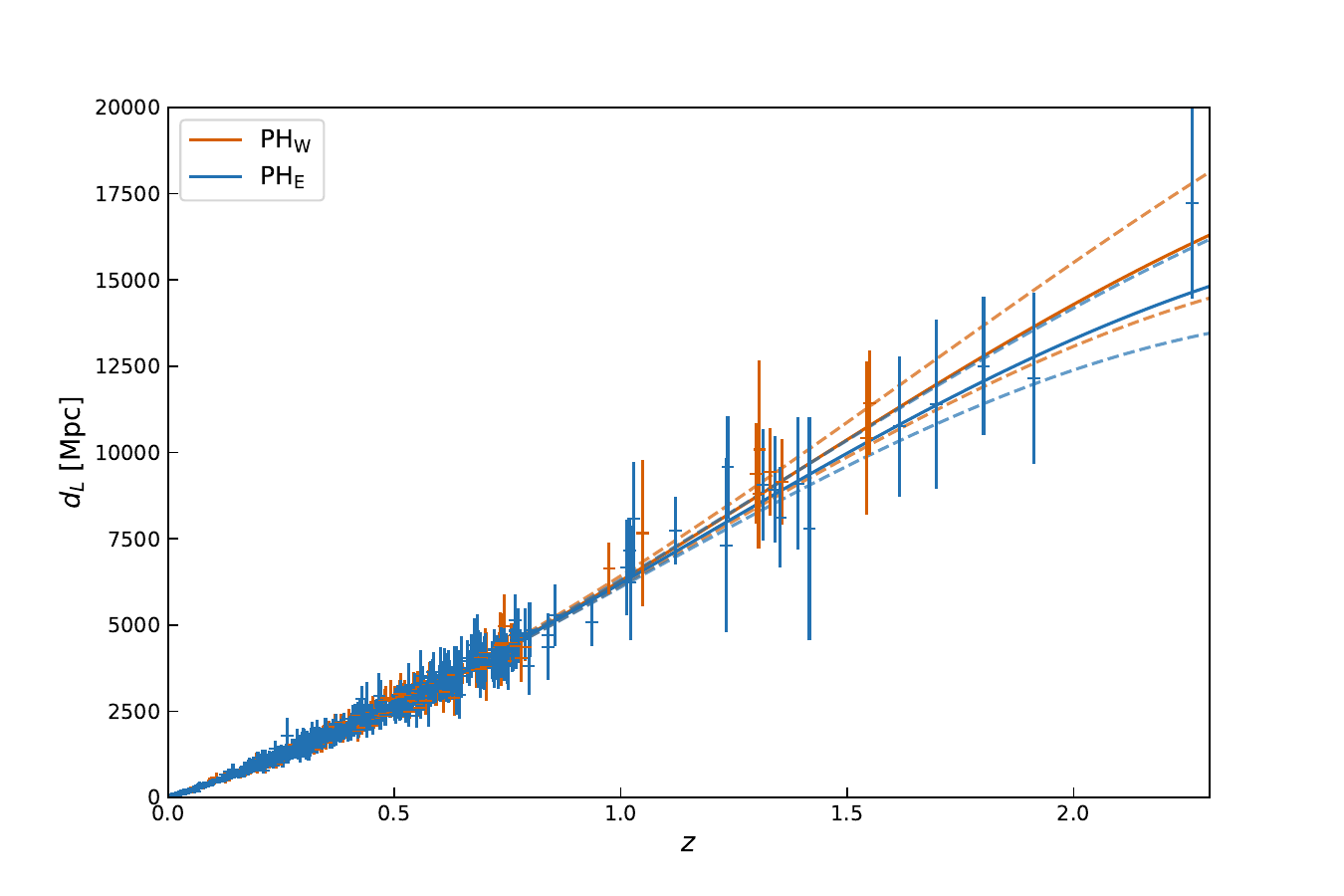}}
\subfloat{\includegraphics[width=0.5\textwidth]{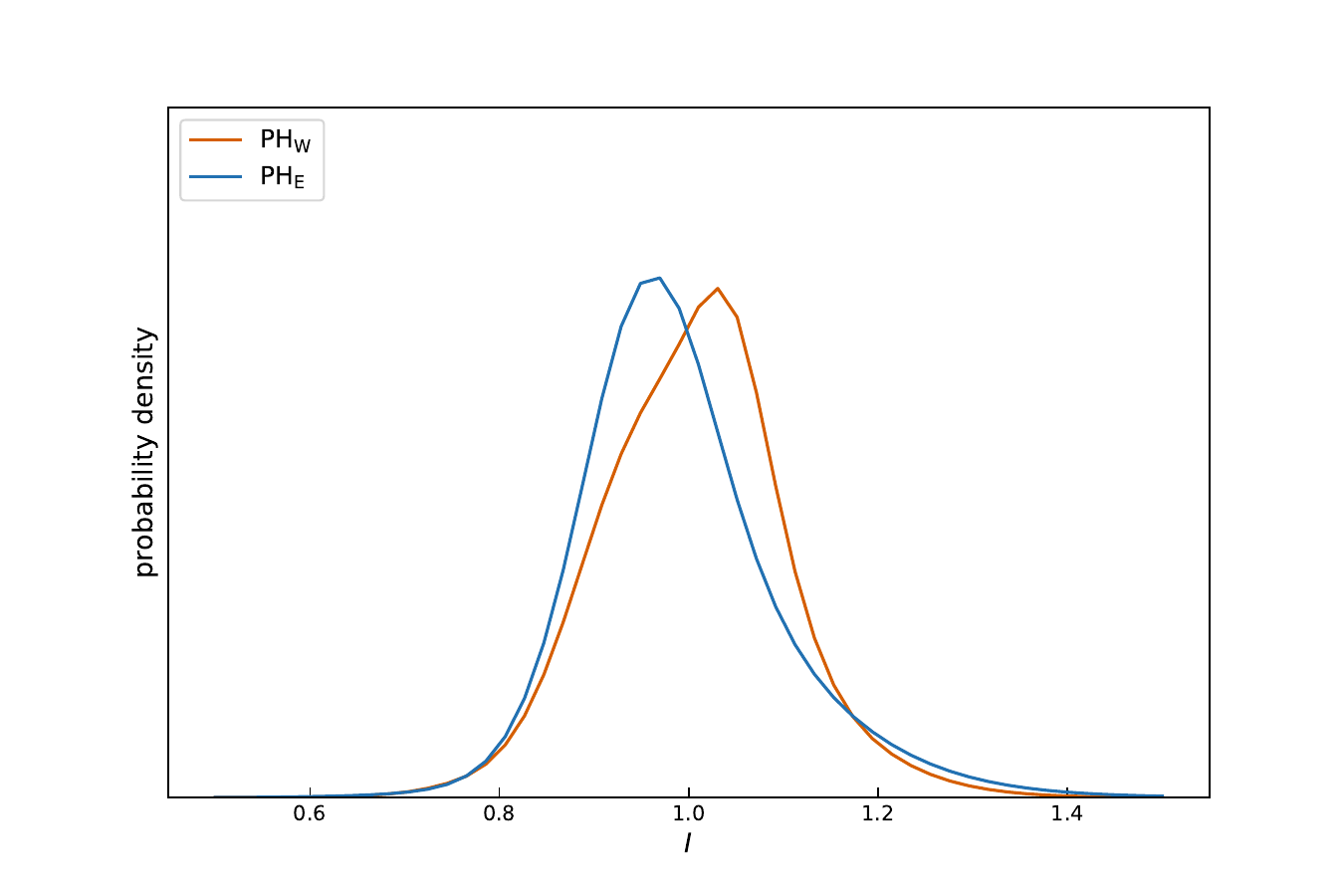}}
\end{center}
\captionsetup{justification=raggedright}
\caption{LEFT: Reconstruction of $d_L(z, \delta_M=0.0478)$ from the $\rm{PH_{W}}$ %dataset (red) 
and $\rm{PH_{E}}$ datasets (red, $\sigma_f=1.2782 \times 10^{4}$~Mpc, $l=2.4030$, and blue, $\sigma_f=1.1363 \times 10^{4}$~Mpc, $l=2.2433$, resp.)% (blue)
, where the bracketing dashed curves define %shaded regions are 
the $68\%$ CL regions of the reconstructions. RIGHT: Constraints on $I$ from the $\rm{PH_{W}}$ %dataset (red) 
and $\rm{PH_{E}}$ datasets (red and blue, resp.)%
.}
\label{fig:we_dl}
\end{figure*}

\begin{figure*}[]
\begin{center}
\subfloat{\includegraphics[width=0.5\textwidth]{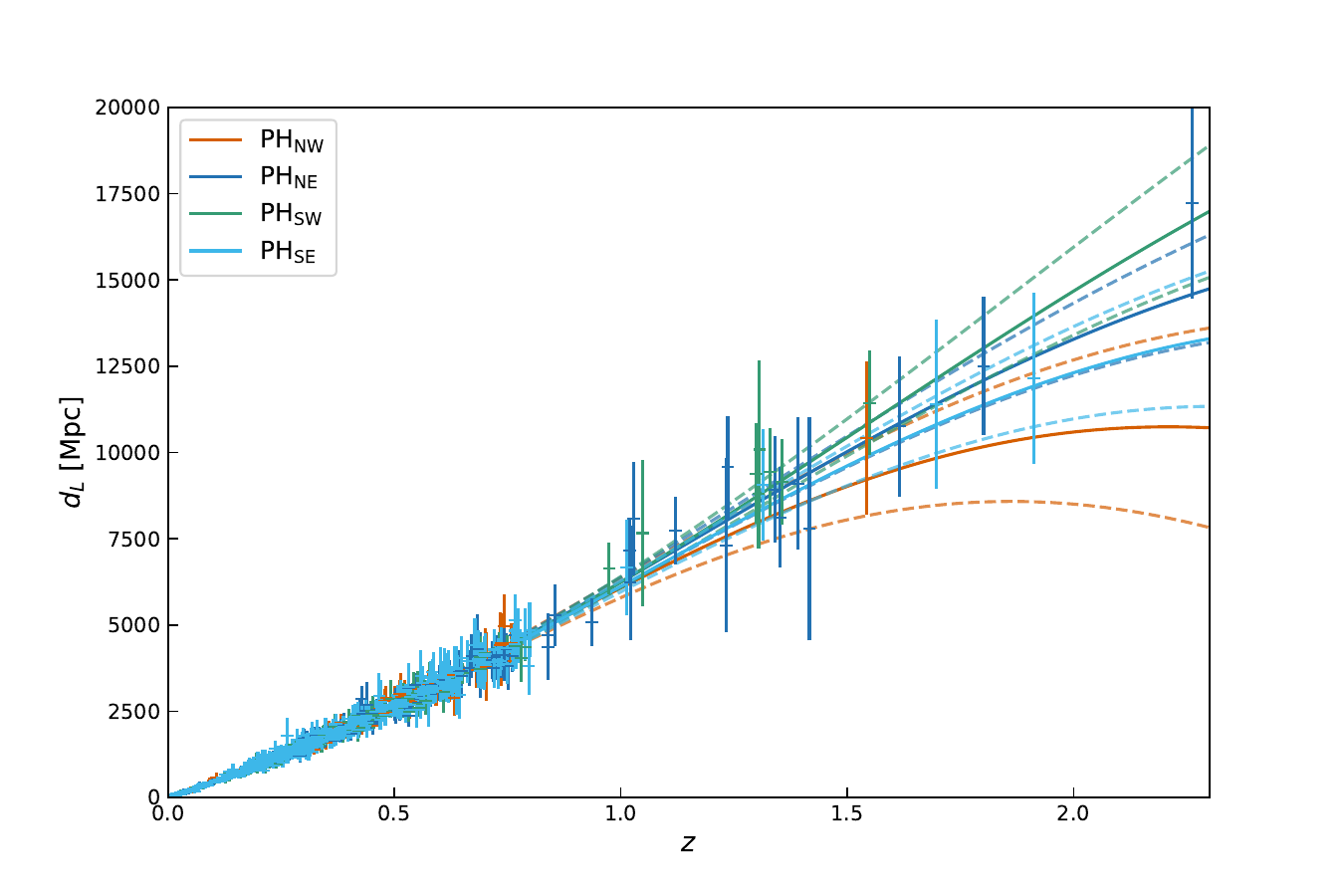}}
\subfloat{\includegraphics[width=0.5\textwidth]{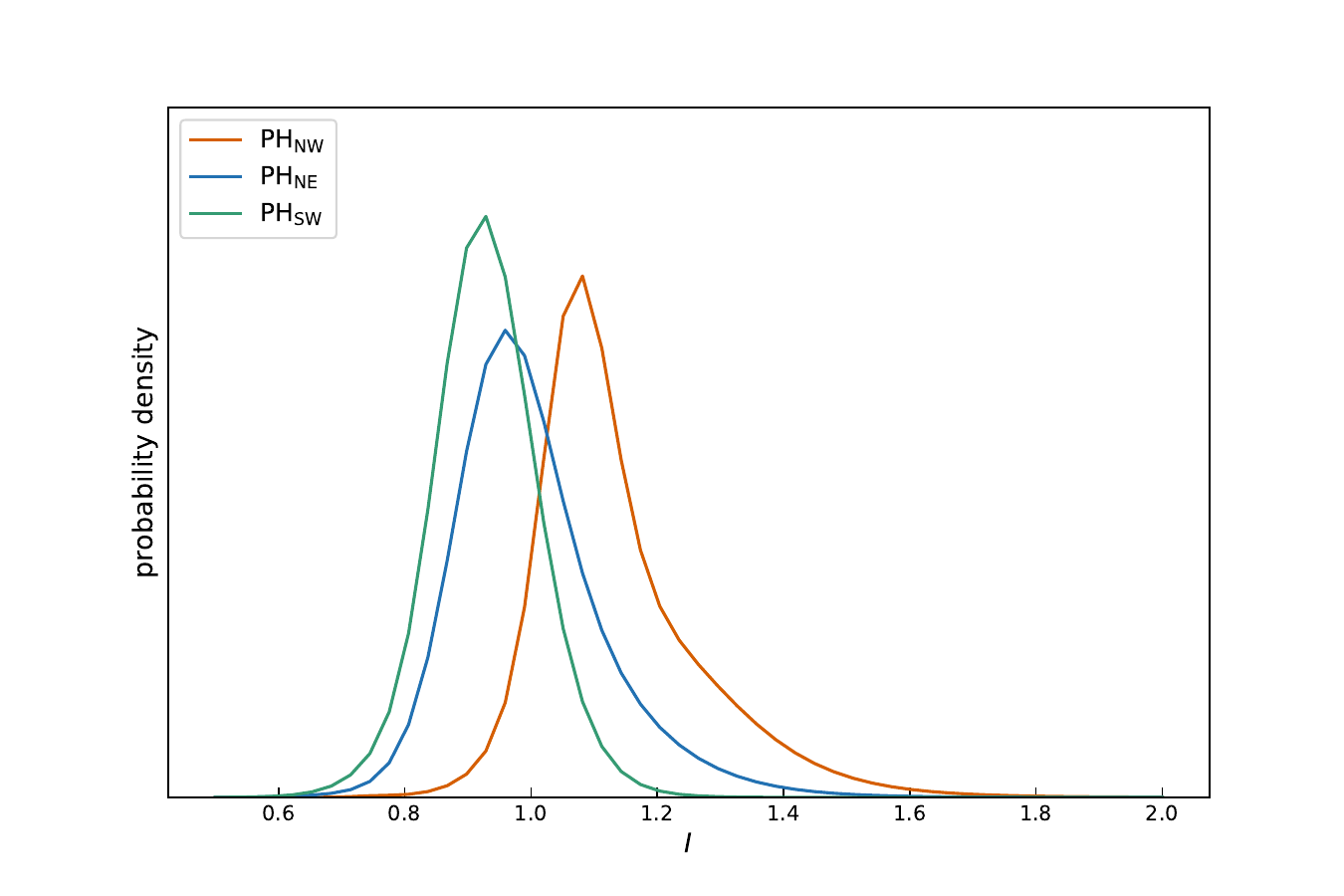}}
\end{center}
\captionsetup{justification=raggedright}
\caption{LEFT: Reconstruction of $d_L(z, \delta_M=0.0478)$ from the $\rm{PH_{NW}}$% dataset (red)
, $\rm{PH_{NE}}$% dataset (blue)
, $\rm{PH_{SW}}$ %dataset (green) 
and $\rm{PH_{SE}}$ datasets (red, $\sigma_f=7.1685 \times 10^{3}$~Mpc, $l=1.6094$, blue, $\sigma_f=1.1006 \times 10^{4}$~Mpc, $l=2.1561$, green, $\sigma_f=1.4135 \times 10^{4}$~Mpc, $l=2.5138$, and cyan, $\sigma_f=8.8715 \times 10^{3}$~Mpc, $l=1.9486$, resp.)% (cyan)
, where the bracketing dashed curves define %shaded regions are 
the $68\%$ CL regions of the reconstructions. RIGHT: Constraints on $I$ from the $\rm{PH_{NW}}$% dataset (red)
, $\rm{PH_{NE}}$ % dataset (blue) 
and $\rm{PH_{SW}}$ datasets (red, blue and green, resp.)%(green)
. Since there is no $d_{\Delta t,i}^H$ in the $\rm{PH_{SE}}$ dataset, there is no constraint on $I$ from it%this dataset
.}
\label{fig:wens_dl}
\end{figure*}

\begin{figure*}[]
\begin{center}
\subfloat{\includegraphics[width=0.5\textwidth]{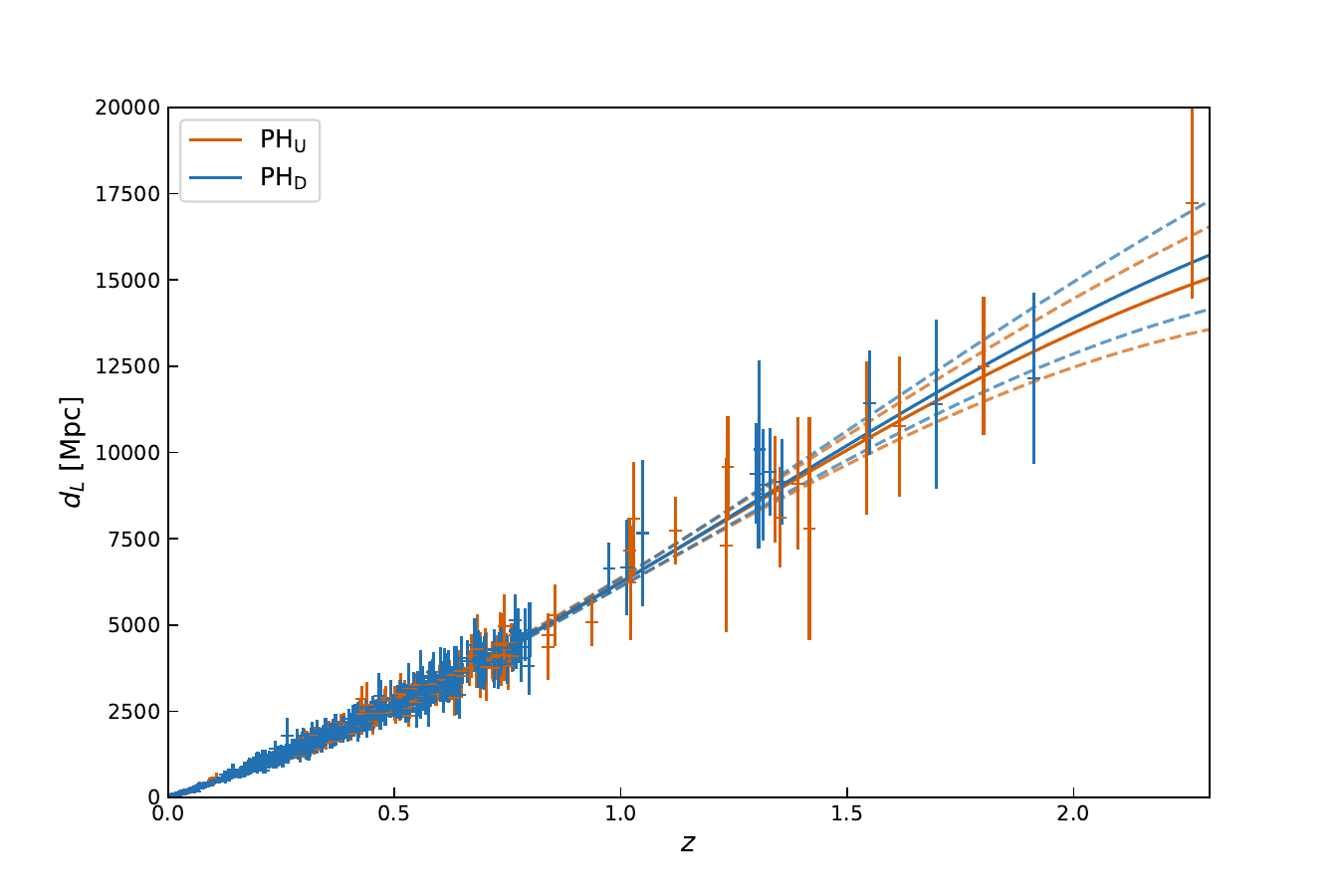}}
\subfloat{\includegraphics[width=0.5\textwidth]{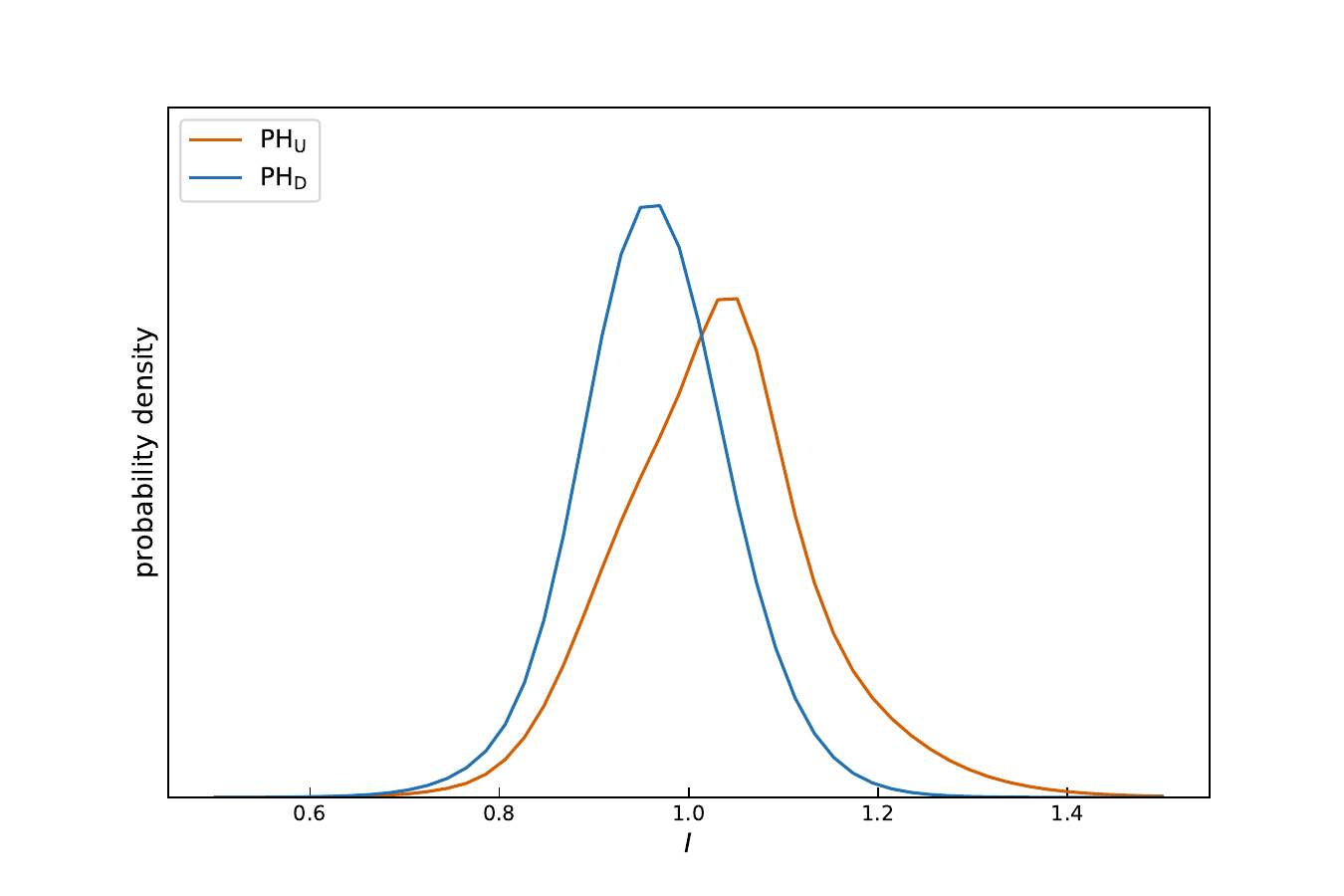}}
\end{center}
\captionsetup{justification=raggedright}
\caption{LEFT: Reconstruction of $d_L(z, \delta_M=0.0478)$ from the $\rm{PH_{U}}$ %dataset (red) 
and $\rm{PH_{D}}$ datasets (red, $\sigma_f=1.1696 \times 10^{4}$~Mpc, $l=2.2704$, and blue, $\sigma_f=1.1314 \times 10^{4}$~Mpc, $l=2.3005$, resp.)% (blue)
, where the %shaded regions are 
the $68\%$ CL regions of the reconstructions. RIGHT: Constraints on $I$ from the $\rm{PH_{W}}$ %dataset (red) 
and $\rm{PH_{D}}$ datasets (red and blue, resp.)% (blue)
.}
\label{fig:ud_dl}
\end{figure*}

Given the absolute magnitude evaluation $M=-19.3+\delta_M=-19.2522^{+0.0594}_{-0.0649}$ (at $68\%$ CL), we can reconstruct $d_L(z)$ from %according to 
Eq.~(\ref{eq:dL}). To probe %the 
cosmic anisotropy, we compare %the 
reconstructions of $d_L(z)$ from every dataset listed in Tab.~\ref{tb:sample}, and display them, such as $\rm{PH_{N}}$ vs. $\rm{PH_{S}}$, % as 
shown in the left panel of Fig.~\ref{fig:ns_dl}, $\rm{PH_{W}}$ vs. $\rm{PH_{E}}$, % as 
shown in the left panel of Fig.~\ref{fig:we_dl}, %or 
$\rm{PH_{NW}}$ vs. $\rm{PH_{NE}}$ vs. $\rm{PH_{SW}}$ vs. $\rm{PH_{SE}}$, % as 
shown in the left panel of Fig.~\ref{fig:wens_dl}, or $\rm{PH_{U}}$ vs. $\rm{PH_{D}}$, shown in the left panel of Fig.~\ref{fig:ud_dl}.
We %can 
find that the $d_L(z)$ reconstructions from $\rm{PH_{N}}$ and $\rm{PH_{S}}$ are consistent with each other, as shown in the left panel of Fig.~\ref{fig:ns_dl}. Similarly, the $d_L(z)$ reconstructions from $\rm{PH_{W}}$ and $\rm{PH_{E}}$, as well as  from $\rm{PH_{U}}$ and $\rm{PH_{D}}$, are also consistent with each other, as shown in the left panel of Fig.~\ref{fig:we_dl} and \ref{fig:ud_dl}, respectively. 
However, the $d_L(z)$ reconstructions from $\rm{PH_{NW}}$ and $\rm{PH_{SW}}$ break consistency %are not very consistent with each other 
at higher redshift at $68\%$ CL, as shown in the left panel of Fig.~\ref{fig:wens_dl}.

To further quantify the %anisotropy 
information about anisotropy carried by $d_L(z)$, we define
\begin{align}
\delta_{i, \rm All}(z) \equiv \frac{d_{L,i}(z)-\overline{d_{L,\rm All}(z)}}{\overline{d_{L,\rm All}(z)}},
\end{align}
where $\overline{d_{L,\rm All}(z)}$ is the mean value of the $d_L(z, \delta_M=0.0478)$ reconstruction from the $\rm{PH_{All}}$ dataset (as plotted as %by 
the blue solid curve in the right panel of Fig.~\ref{fig:all_dl}) and $d_{L,i}(z)$ is the reconstruction of $d_L(z, \delta_M=0.0478)$ from the other datasets listed in Tab.~\ref{tb:sample}.
We list the constraints on $\delta_{i, \rm All}(z=0.5)$, $\delta_{i, \rm All}(z=1.0)$, $\delta_{i, \rm All}(z=1.5)$ and $\delta_{i, \rm All}(z=2,0)$ at $68\%$ CL. in Tab.~\ref{tb:dL}.
\begin{table*}
\captionsetup{justification=raggedright}
\caption{Constraints on $\delta_{i, \rm All}(z=0.5)$, $\delta_{i, \rm All}(z=1.0)$, $\delta_{i, \rm All}(z=1.5)$ and $\delta_{i, \rm All}(z=2,0)$ at $68\%$ CL.}
\label{tb:dL}
\begin{tabular}{ccccc}
\hline
\hline
$\delta_{i, \rm All}(z)$& $z=0.5$ &  $z=1.0$ & $z=1.5$ & $z=2.0$\\
\hline
$\delta_{\rm N,All}(z)$&  $0.0049\pm0.0096$ &  $0.0008\pm0.0203$ & $-0.0130\pm0.0412$ & $-0.0338\pm0.0709$ \\
$\delta_{\rm S,All}(z)$&  $-0.0032\pm0.0075$ &  $-0.0016\pm0.0187$ & $0.0024\pm0.0414$ & $0.0049\pm0.0744$ \\
\hline
$\delta_{\rm W,All}(z)$&  $-0.0005\pm0.0099$ &  $0.0065\pm0.0229$ & $0.0155\pm0.0493$ & $0.0238\pm0.0871$   \\
$\delta_{\rm E,All}(z)$&  $0.0003\pm0.0073$  &  $-0.0055\pm0.0174$ & $-0.0229\pm0.0366$ & $-0.0481\pm0.0644$  \\
\hline
$\delta_{\rm NW,All}(z)$&  $0.0144\pm0.0163$ &  $-0.0269\pm0.0454$ & $-0.1172\pm0.0948$ & $\bf -0.2412\pm0.1498$ \\
$\delta_{\rm NE,All}(z)$&  $0.0021\pm0.0114$ &  $0.0001\pm0.0228$  & $-0.0185\pm0.0442$ & $-0.0485\pm0.0750$ \\
$\delta_{\rm SW,All}(z)$&  $-0.0101\pm0.0124$ &  $-0.0003\pm0.0263$  & $0.0220\pm0.0524$ & $\bf 0.0515\pm0.0914$   \\
$\delta_{\rm SE,All}(z)$&  $-0.0016\pm0.0091$  &  $-0.0194\pm0.0256$   & $-0.0588\pm0.0561$ & $-0.1181\pm0.0959$  \\
\hline
$\delta_{\rm U,All}(z)$&  $0.0052\pm0.0096$ &  $0.0007\pm0.0204$ & $-0.0139\pm0.0413$ & $-0.0354\pm0.0713$   \\
$\delta_{\rm D,All}(z)$&  $-0.0027\pm0.0075$  &  $-0.0016\pm0.0188$ & $-0.0008\pm0.0417$ & $-0.0041\pm0.0746$   \\
\hline 
\hline
\end{tabular}
\end{table*}
We find that $\delta_{\rm NW,All}(z=2.0)=-0.2412 \pm 0.1498$ (at $68\%$ CL) and $\delta_{\rm SW,All}(z=2.0)=0.0515 \pm 0.0914$ (at $68\%$ CL) are mutually inconsistent%not very consistent with each other
.
This %over $1\sigma$ 
inconsistency of more than $1\sigma$ can be regarded as a weak preference for %the 
cosmic anisotropy.
% \Mout{Since we reconstruct $d_L(z)$ directly, the anisotropic information maybe carried by the Hubble parameter $H(z)$ or \Mov{by }the cosmic curvature $\Omega_k$\Mov{, which can be\Mcom{???} }% has been 
% included as $d_L(z)=d_L(z,H(z),\Omega_k)$ even thought there is no specific cosmological model %to be 
% assumed.}
However, as the $d_L(z)$ reconstruction is model independent, it takes into account any possible source of anisotropy, including the Hubble parameter $H(z)$ or cosmic curvature $\Omega_k$. The anisotropy effect can be expressed in $d_L(z)=d_L(z,H(z),\Omega_k)$ even thought there is no specific cosmological model 
assumed.
Therefore, we can not disentangle the cosmic anisotropy source between Hubble expansion %\Mout{tell the source of this cosmic anisotropy is the dark energy or}
and %the 
cosmic curvature.

%In fact, the above inconsistency between $d_L(z)$ reconstructions can also result from the inconsistency between data or the paucity of data.
In fact, the above %over $1\sigma$ 
inconsistency between $d_L(z)$ from the $\rm{PH_{W}}$ and $\rm{PH_{E}}$ datasets can either %can 
be regarded as %the 
anisotropic information from the Pantheon+ sample~\cite{Brout:2022vxf} or as %the 
inconsistency in %between 
SNIa data from %itself due to the 
potential systematic uncertainties.
Undoubtedly, if the above %over $1\sigma$ 
inconsistency %between $d_L(z)$ reconstructions 
do result from %the 
cosmic anisotropy, the anisotropic information should affect %carry by 
both the Pantheon+ sample~\cite{Brout:2022vxf} and the H0LiCOW sample~\cite{Wong:2019kwg}. 
We t%T
herefore %, we 
also need to check whether this anisotropic information is present in %with 
the H0LiCOW sample~\cite{Wong:2019kwg}.
Here, we use the prediction of $d_{\Delta t}$ from different dataset to test their %do the 
consistency %test 
and probe the impact of potential systematic uncertainties% in data
. 
Firstly, we define the average derived--to--measured effective time--delay distance
\begin{align}
\label{eq:I}
I\equiv  \frac{1}{N}\sum_{i=1}^{N}  \frac{d_{\Delta t,i}^P}{d_{\Delta t,i}^H},
\end{align}
where $d_{\Delta t,i}^H$ is the measured effective time-delay distance of the H0LiCOW $i$-th lens system %of a certain dataset 
listed in Tab.~\ref{tb:sample} and $d_{\Delta t,i}^P$ is that %the 
derived %one 
from the partial Pantheon+ sample restricted to %from 
the same dataset, using Eq.~\eqref{eq:Dt}. 
We compare the constraints on $I$ from every dataset listed in Tab.~\ref{tb:sample}, and represent them separately: %such as 
$\rm{PH_{N}}$ vs. $\rm{PH_{S}}$ is %as 
shown in %the right panel of 
Fig.~\ref{fig:ns_dl}, right panel, $\rm{PH_{W}}$ vs. $\rm{PH_{E}}$ %as shown 
in %the right panel of 
Fig.~\ref{fig:we_dl}'s right panel, % or 
$\rm{PH_{NW}}$ vs. $\rm{PH_{NE}}$ vs. $\rm{PH_{SW}}$ is portrayed in %as shown in the right panel of 
Fig.~\ref{fig:wens_dl}'s right panel, while $\rm{PH_{U}}$ vs. $\rm{PH_{D}}$ is displayed in Fig.~\ref{fig:ud_dl}'s right panel. We also list the constraints on $I$ at $68\%$ CL in Tab.~\ref{tb:I}.
We %can 
find that the tested
$I$s are consistent pairwise: %from 
$\rm{PH_{N}}$ and %is consistent with $I$ from 
$\rm{PH_{S}}$; %$I$ from 
$\rm{PH_{W}}$ and %is consistent with $I$ from 
$\rm{PH_{E}}$; %$I$ from 
$\rm{PH_{NE}}$ and %is consistent with $I$ from 
$\rm{PH_{SW}}$; $\rm{PH_{U}}$ and 
$\rm{PH_{D}}$; they are all %of them are also 
consistent with $I=1$, marking agreement between H0LiCOW and Pantheon+ data.
Although the pairs of samples from $\rm{PH_{U}}$ and $\rm{PH_{N}}$ on one hand, and from $\rm{PH_{D}}$ and $\rm{PH_{S}}$ on another hand, contain populations of SNIa with different sizes, they include the same lensed systems.
The values for $\delta_{\rm U,All}(z)$ and $\delta_{\rm D,All}(z)$ are both consistent with 0 and their $I$s are also consistent with each other, which mark no detectable difference between these two hemispheres in effective time-delay distance or luminosity distance, parallel to the CMB dipole. 
Note that as %(Since 
there is no lensed quasar in the %$d_{\Delta t,i}^H$ in 
$\rm{PH_{SE}}$ dataset, there is no constraint on $I$ from a $d_{\Delta t,i}^H$ for this dataset. %)
However, we obtain %But 
$I=1.1051^{+0.1427}_{-0.0813}$ (at $68\%$ CL) from $\rm{PH_{NW}}$, which deviates at more than $1\sigma$ % is deviate 
from $I=1$. Thus, % at $68\%$ CL, which means 
the prediction from %by 
the partial Pantheon+ sample is not %very 
consistent with the corresponding lensed system measurements for $\rm{PH_{NW}}$, % by lens systems for $\rm{PH_{NW}}$ dataset.
which means that %That is to say, 
the $\rm{PH_{NW}}$ dataset cannot be considered %is not very 
self-consistent.
Therefore, we cannot %can't 
guarantee that the reconstruction inconsistencies in %over $1\sigma$ inconsistency between 
$d_L(z)$ between %reconstructions from 
$\rm{PH_{NW}}$ and $\rm{PH_{SW}}$ result %s 
from %the 
cosmic anisotropy\Mov{ rather than SNIa data systematics}.
%However, the preference for the cosmic anisotropy is so weak that we should confirm it further by other methods. 


\begin{table}
\captionsetup{justification=raggedright}
\caption{Constraints on $I$ at $68\%$ CL. Since there is no $d_{\Delta t,i}^H$ in the $\rm{PH_{SE}}$ dataset, there is no constraint on $I$.}
\label{tb:I}
\begin{tabular}{cc}
\hline
\hline
Dataset & $I$ \\
\hline
$\rm{PH_{N}}$ &  $1.0279^{+0.0974}_{-0.1030}$   \\
$\rm{PH_{S}}$ & $0.9583^{+0.0789}_{-0.0774}$  \\
\hline
$\rm{PH_{W}}$  &  $1.0068^{+0.0882}_{-0.0989}$ \\
$\rm{PH_{E}}$  &  $0.9783^{+0.1093}_{-0.0844}$   \\
\hline
$\rm{PH_{NW}}$  &   $1.1051^{+0.1427}_{-0.0813}$ \\
$\rm{PH_{NE}}$  &  $0.9820^{+0.1232}_{-0.0923}$   \\
$\rm{PH_{SW}}$  &  $0.9273^{+0.0826}_{-0.0787}$ \\
$\rm{PH_{SE}}$  &  $-$   \\
\hline
$\rm{PH_{U}}$  &  $1.0289^{+0.0975}_{-0.1027}$ \\
$\rm{PH_{D}}$  &  $0.9630^{+0.0804}_{-0.0773}$   \\
\hline 
\hline
\end{tabular}
\end{table}

\section{Summary and Discussion}
\label{sec:SD}
In this paper, we reconstruct the luminosity distance %try to use 
$d_L(z)$ model-independently %reconstruction 
from different spatially selected datasets to probe %the 
cosmic anisotropy. We start with %At the beginning, however, we have 
no prior knowledge of $M$, the absolute magnitude of SNIas, into which $H_0$ %which 
is %also 
degenerate% with $H_0$
. We eliminate that degeneracy by focusing on %Therefore, we turn to Eq.~(\ref{eq:m2DLt}) to isolate the unknown information of $H_0$ and then do 
the reconstruction of $\tilde{D}_L(z)$, the dimensionless luminosity distance at a fiducial $H_0$ that only depends on $M$ through a parameter $\delta_M$ in Eq.~\eqref{eq:m2DLt}, from the whole Pantheon+ sample~\cite{Brout:2022vxf}% according to this equation with a parameter $\delta_M$
. In addition, %By 
anchoring the $\tilde{D}_L(z)$ reconstruction with the %whole 
H0LiCOW sample~\cite{Wong:2019kwg}, we can extract a measurement %the information 
of $M$ as %can be obtained, 
$M=-19.2522^{+0.0594}_{-0.0649}$ (at $68\%$ CL). We f%F
inally reintroduce %, we incorporate the unknown information of 
$H_0$ and $\Omega_k$, the unknown value of the curvature parameter, into a model-independent %$d_L(z)$ and do the 
reconstruction of $d_L(z)$ from the different datasets listed in Tab.~\ref{tb:sample}, using % according to  
Eq.~\eqref{eq:dL}. We find that our reconstructions of $d_L(z)$ %reconstructions 
from these datasets, covering % making up 
the complete Pantheon+ sample, show consistency between most of the subsamples, % are almost consistent with each other, 
as shown in the left panel of Fig.~\ref{fig:ns_dl}, Fig.~\ref{fig:we_dl}, %and 
Fig.~\ref{fig:wens_dl} and Fig.~\ref{fig:ud_dl}, except between $\rm{PH_{NW}}$ and $\rm{PH_{SW}}$, at %that $d_L(z)$ reconstructions from $\rm{PH_{NW}}$ and $\rm{PH_{SW}}$ are not very consistent with each other at 
higher redshift (at $68\%$ CL), as shown in the left panel of Fig.~\ref{fig:wens_dl}.
\Mov{%As listed in Tab.~\ref{tb:dL}, the prediction of $d_L(z=0.5)$ (resp. $d_L(z=2.0)$) from $\rm{PH_{SW}}$ is smaller (resp. larger) than its counterparts from the other subsets, which is consistent with the preferred directions of cosmic anisotropy in the SNIa sky ($l\sim310$°$,b\sim-15$°) where the maximum $H_0$ and the minimum $\Omega_m$ are detected~\cite{Hu:2023eyf,Hu:2024qnx}.-----------------
Our results are consistent with those of~\cite{Hu:2023eyf,Hu:2024qnx}. The low (resp. high) redshift predictions listed in Tab.~\ref{tb:dL} for $d_L(z=0.5)$ (resp. $d_L(z=2.0)$) from $\rm{PH_{SW}}$ are smaller (resp. larger) than their counterparts from the other subsets.
This is in agreement with the~\cite{Hu:2023eyf,Hu:2024qnx} detections of a maximum $H_0$ and a minimum $\Omega_m$ in the ($l\sim310$°$,b\sim-15$°) direction of the SNIa sky, as it corresponds to an anisotropy in the SW sky of our division and follows the low (resp. high) redshift relation between luminosity distance and $H_0$ (resp. $\Omega_m$).
%
% $d_L(z=2.0)$ is larger = minimum $\Omega_m$
% $d_L(z=0.5)$ is smaller =  maximum $H_0$
% integration of 1/Ome_m is sum of 1/Ome_m, so which contribute more than 1/H0 for higher red redshift
% comparison is between different subset?
% so $d_L(z=2.0)$ is larger = minimum $\Omega_m$
% 1/Ome_m(1+z)^3 is more sensitive to the difference between different subsets than 1/H_0 and 1/Ome_Lambda for high red redshift because (1+z)^3 is amplify this difference.
% ----------
% This is in agreement with the fact that the detections of a maximum $H_0$ and a minimum $\Omega_m$ in the SNIa sky are around ($l\sim310$°$,b\sim-15$°)~\cite{Hu:2023eyf,Hu:2024qnx} which corresponds to an anisotropy in the SW sky of our division.
% ----------
% This is in agreement with the~\cite{Hu:2023eyf,Hu:2024qnx} detections of the ($l\sim310$°$,b\sim-15$°) direction of the SNIa sky, for which $H_0$ (resp. $\Omega_m$) is found maximum (resp. minimum), as this direction corresponds to an anisotropy in the SW sky of our division.
}
After a further consistency test comparing the Pantheon+ and H0LiCOW samples, we find that the $\rm{PH_{NW}}$ dataset is not %very 
self-consistent.
Therefore, %the 
cosmic anisotropy is just one possible %of 
source for the above mentioned more than %s of the over 
$1\sigma$ level inconsistency between %$d_L(z)$ reconstructions from 
$\rm{PH_{NW}}$ and $\rm{PH_{SW}}$.

The method applied in our paper can not constrain some important cosmological parameters, such as $H_0$ or $\Omega_k$. While %Even though 
they can carry %the 
anisotropic information and the constraints on $H_0$ can help deal with the Hubble tension, %the 
direct reconstruction of $d_L(z)$  is unable to distinguish, in the end, between Hubble parameter and curvature sources of anisotropy%has given up these advantages
. Of course, %the 
direct reconstruction of $H(z)$ retains merit on its own, such as giving %is a good idea which can further give 
constraints on $H_0$ or $\Omega_k$ as was done %did 
in~\cite{Liao:2019qoc}. However, since our method departs from %has given up 
the fiducial relation $M_*=-19.3-5\log_{10}[70/(3\times10^{10})]$, % and 
the presence of two new parameters $\delta_M$ and $\delta_{70}$ makes %the 
direct reconstruction of $H(z)$ impossible. Should one insist in constraining cosmological parameters, one can either turn to the reconstruction of $d_L(z)\approx\frac{cz}{H_0}$ at lower redshift, that only allows to circumscribe $H_0$ from SNIa with lower redshifts, or to the reconstruction of $d_L(z)=\frac{c(1+z)}{H_0}\int_0^z\frac{dz'}{(\Omega_m(1+z')^3+\Omega_\Lambda)^{1/2}}$ at higher redshift, to constrain $\{H_0, \Omega_m,\Omega_\Lambda\}$ in the $\Lambda$CDM model. In other words, our model-independent method is more general and powerful than the methods from~\cite{Deng:2018yhb,Sun:2018cha,Deng:2018jrp,Zhao:2019azy,Hu:2023eyf,Zhao:2013yaa,Lin:2015rza,Chang:2019utc,Wang:2023reg} %described above 
and our results can be applied to any cosmological models. 

Although our paper is only using six lensed quasars from H0LiCOW, they constitute a large enough sample %there are only six lenses applied in our paper, the number of them is large enough 
to serve as good anchors. More precisely, according to Eq.~\eqref{eq:LdM}, each lensed system's %the 
effective time-delay distance $d_{\Delta t}$ %of every lens system are 
is combined to anchor $\delta_M$. Of course, the number of lensed system %s 
may affects the constraints on $\delta_M$. As hinted in the right panel of Fig.~\ref{fig:all_dl}, however, the method's shifted %a different 
mean value of $\delta_M\neq0.0478$ (or $M\neq19.2522$) used in the reconstruction of %to reconstruct 
$d_L(z)$ according to Eq.~\eqref{eq:dL} just shifts the results shown in the left panel of Fig.~\ref{fig:ns_dl}, Fig.~\ref{fig:we_dl}, %and 
Fig.~\ref{fig:wens_dl} and Fig.~\ref{fig:ud_dl} but does not change their consistency. Therefore, the lensed sample size is less important than %only thing we should care about is not the number of lenses but 
the consistency between them and SNIa data, which is what we have checked in % which we have done with Eq.~(
Eq.~\eqref{eq:I}.

\begin{acknowledgments}
%Ke Wang is supported by grants from NSFC (grant No.12247101). 
MLeD acknowledges the financial support by the Lanzhou University starting
fund, the Fundamental Research Funds for the Central Universities
(Grant No. lzujbky-2019-25) and NSFC (grant No.12247101).
\end{acknowledgments}

\end{document}